\documentclass[11pt]{article}
\usepackage{latexsym,cite,amssymb,amsmath}

\usepackage{times}
\usepackage{hyperref} 
\makeatletter
\@addtoreset{equation}{section} \makeatother

\input amssym.def
\input amssym.tex

\newcommand{\half}{{{\textstyle\frac{1}{2}}}}

\newcommand{\be}{\begin{equation}}
\newcommand{\ee}{\end{equation} }
\newcommand{\beqa}{\begin{eqnarray} }
\newcommand{\eeqa}{\end{eqnarray} }
\newcommand{\ba}{\begin{array}}
\newcommand{\ea}{\end{array}}

\newcommand{\SODD}{\mathbf{SO}(D,D)}

\newcommand{\ODD}{\mathbf{O}(D,D)}
\newcommand{\sodd}{\mathbf{so}(D,D)}

\newcommand\rd{{\rm d}}

\newcommand\cH{{\cal H}}

\newcommand\cK{{\cal K}}

\newcommand\cO{{\cal O}}

\newcommand\cR{{\cal R}}

\newcommand\cT{{\cal T}}

\newcommand\hcL{{\hat{\cal L}}}
\newcommand\tcL{{\widetilde{\cal L}}}
\newcommand\wT{{\widetilde{T}}}
\newcommand\nD{{D^{\prime}\!}}
\newcommand\nDGamma{{\Gamma^{\prime}\!}}

\newcommand\rbC{{\scriptscriptstyle{\mathbf{C}}}}

\newcommand\dis{\displaystyle}

\def\tx{\tilde{x}}

\def\Tw{{T_{\omega}}}

\def\bP{{\bar{P}}}

\begin{document}
\begin{titlepage}
\title{%\vskip -60pt
%\vskip 20pt
\vskip 2cm
Differential geometry with a projection:\\Application to  double field theory\\~\\}
\author{\sc Imtak Jeon,${}^{\dagger}$  \mbox{~~\,\,}
Kanghoon Lee${}^{\sharp}$ \mbox{~~\,\,}and\mbox{\,\,~~} Jeong-Hyuck Park${}^{\dagger}$}
\date{}
\maketitle \vspace{-1.0cm}
\begin{center}
~~~\\
${}^{\sharp}$Center for Quantum Spacetime, Sogang University, Shinsu-dong, Mapo-gu, Seoul 121-742, Korea\\
%\texttt{kanghoon@sogang.ac.kr}
%\texttt{}\\
~{}\\
${}^{\dagger}$Department of Physics, Sogang University, Shinsu-dong, Mapo-gu, Seoul 121-742, Korea\\
%\texttt{imtak@sogang.ac.kr    ~~~~~park@sogang.ac.kr}
~{}\\
%{\small{{Electronic correspondence:} \texttt{{{{~imtak,kanghoon,park@sogang.ac.kr}}}}}}\\
{\small{{E-mail:} \texttt{~{{{imtak@sogang.ac.kr~~kanghoon@sogang.ac.kr~~park@sogang.ac.kr}}}}}}
~~~\\
\end{center}
\begin{abstract}
\vskip0.5cm
\noindent
In recent development  of double field theory, as  for the description of   the massless sector of closed strings,    the spacetime dimension is formally  doubled, \textit{i.e.~}from $D$ to $D{+D}$,   and    the T-duality is realized manifestly  as a global $\ODD$ rotation.  In this paper, we conceive    a differential geometry characterized by a $\ODD$ symmetric projection, as the underlying mathematical structure of  double field theory. We introduce     a 
 differential operator compatible with the projection,  which, contracted  with the projection, can be covariantized and may replace the ordinary derivatives in the generalized Lie derivative that generates  the gauge symmetry of double field theory.    We construct  various gauge covariant tensors  which include   a scalar and  a tensor carrying   two $\ODD$ vector indices. 
 \end{abstract}

%%%
{\small
\begin{flushleft}
~~~~~~~~\textit{PACS}: 11.25.-w\\
%%Strings and branes \\
~~~~~~~~\textit{Keywords}: Double field theory, T-duality, Differential geometry.
\end{flushleft}}
\thispagestyle{empty}
%%%%
\end{titlepage}
\newpage

\tableofcontents %%
%%\begin{document} --> JHEP
%%%
%%%%%%%%%%%%%%%%%%%%%%%%%%%%%%%%%%%%%%%%%%%%%%%%%%%%%%%%%%%%%%%%%%%%%%%%%%%%%%%%%%%%%%%%%%%%%%%%
%%%%%%%%%%%%%%%%%%%%%%%%%%%%%%%%%%%%%%%%%%%%%%%%%%%%%%%%%%%%%%%%%%%%%%%%%%%%%%%%%%%%%%%%%%%%%%%%
\section{Introduction}
One of  the  characteristic features of string theory, in contrast to ordinary particle physics, is the existence  of  T-duality. When closed strings  wrap around a torus,  they can acquire winding modes, $w$, in addition to the momentum mode, $p$, such that their left and right moving modes have the expansion (see \textit{e.g.}~\cite{Becker:2007zj}):
\be
\ba{l}
X^{\mu}_{L}(\sigma^{+})=\half(x^{\mu}+\tx^{\mu})+\half(p^{\mu}+w^{\mu})\sigma^{+}+
\cdots\,,\\
X^{\mu}_{R}(\sigma^{-})=\half(x^{\mu}-\tx^{\mu})+\half(p^{\mu}-w^{\mu})\sigma^{-}+
\cdots\,,\\
\ea
\ee
where $\sigma^{\pm}=\tau\pm\sigma$ are the usual light-cone coordinates of the string worldsheet and  the ellipses denote the harmonic modes.  As the left and right moving modes are independent (up to the level matching condition), there are  independent left and right center of mass positions, $x\pm\tx$, and also  momenta, $p\pm w$. Under T-duality, the  left and right moving modes transform as~\cite{Giveon:1994fu},
\be
X^{\mu}_{L}+X^{\mu}_{R}~\longrightarrow~X^{\mu}_{L}-X^{\mu}_{R}\,,
\ee
such that the two pairs, $(x,p)$ and $(\tx,w)$, are exchanged by each other,
\be
(x,\tx,p,w)~\longrightarrow~(\tx,x,w,p)\,.
\ee
Namely, before T-duality, $\tx$ is absent and only $x$ appears as  the total center of mass position, yet  after T-duality we encounter the opposite situation. T-duality is an exact  symmetry  working  not only on a  flat spacetime background for the string energy spectrum  but also on a generic  background for the full interaction. \\

\noindent The low energy effective action for the closed string massless sector is of the well-known form: 
\be
\dis{S=\int\rd x^{D}\sqrt{-g}e^{-2\phi}\left[\,R_{g}+4(\partial \phi)^{2}-\textstyle{\frac{1}{12}}H^{2}\,\right]\,,}
\label{NSaction}
\ee
where $g_{\mu\nu}=g_{\nu\mu}$ is the  $D$-dimensional   spacetime metric with its scalar curvature, $R_{g}$;    $\phi$ is the  string theory dilaton; and $H$ is the three form field strength of a two form gauge field, $B_{\mu\nu}=-B_{\nu\mu}$.  An amazing  property of   the action (\ref{NSaction}) which was first  noted by Buscher~\cite{Buscher:1985kb,Buscher:1987sk,Buscher:1987qj} is that, it realizes the  T-duality as  nonlinear symmetry transformations of the fields.   If we introduce a $2D\times 2D$ matrix with the indices of the doubled  spacetime, $A,B=1,2,\cdots,2D$, 
\be
\cH_{AB}=\left(\ba{cc}
g^{\mu\nu}&-g^{\mu\kappa}B_{\kappa\sigma}\\
B_{\rho\kappa}g^{\kappa\nu}&~~g_{\rho\sigma}-B_{\rho\kappa}g^{\kappa\lambda}B_{\lambda\sigma}
\ea
\right)\,,
\label{gH}
\ee
the T-duality transformation rule can be summarized in a compact  manner~\cite{Giveon:1988tt,Tseytlin:1990nb,Tseytlin:1990va,Siegel:1993th,Siegel:1993xq}:
\be
\ba{ll}
\cH~\longrightarrow~\cT\cH\cT\,,~~~&~~~~\phi~\longrightarrow~\phi 
-\half\ln\det(g+B)\,,
\ea
\label{TdualityTr}
\ee
where $\cT$ is a $2D\times 2D$ constant  matrix given by
\be
\cT:=\left(\ba{cc}0&1\\1&0\ea\right)\,.
\label{TODD}
\ee
This matrix also serves   the $\ODD$ invariant metric which we denote separately by $\eta_{AB}$,
\be
\eta:={\textstyle{\left(\ba{cc}0&1\\1&0\ea\right)}}\,.
\label{ODDeta}
\ee
Since $\cT^{t}\eta\cT=\eta$, {\,}the T-duality matrix (\ref{TODD}) is identified as a $\ODD$ rotation.\\

\noindent A recent remarkable advance on T-duality was made by  Hohm, Hull and Zwiebach
 in Ref.\cite{Hohm:2010pp}, based on their earlier works~\cite{Hull:2009mi,Hull:2009zb,Hohm:2010jy}. They  called $\cH_{AB}$   (\ref{gH})  ``generalized metric" and  constructed a double field theory action for  it,  on the  doubled  spacetime having  coordinates,  $y^{A}=(x^{\mu},\tx^{\nu})$,  
\be
\dis{S_{\scriptstyle{\rm{DFT}}}=\int\rd y^{2D}~e^{-2d}\,\cR(\cH,d)\,.}
\label{DFTS}
\ee
With  a double field theory `dilaton', $d$, 
\be
e^{-2d}=\sqrt{-g}e^{-2\phi}\,,
\label{dphi}
\ee
the scalar  Lagrangian, $\cR(\cH,d)$,  which was called ``generalized scalar curvature" in \cite{Hohm:2010pp},  is given by
\be
\ba{ll}\cR(\cH,d)=&
\cH^{AB}\left(4\partial_{A}\partial_{B}d-4\partial_{A}d\partial_{B}d+\textstyle{\frac{1}{8}}
\partial_{A}\cH^{CD}\partial_{B}\cH_{CD}
-\half\partial_{A}\cH^{CD}\partial_{C}\cH_{BD}\right)\\
{}&~+4\partial_{A}\cH^{AB}\partial_{B}d-\partial_{A}\partial_{B}\cH^{AB}\,.
\ea
\label{cRODD}
\ee
All the spacetime indices, $A,B,C,\cdots,$ are $\ODD$ vector indices which can be  raised or lowered by the constant   $\ODD$ invariant metric, $\eta$ in  (\ref{ODDeta}).

As a  field theory counterpart of the level matching condition in the closed string theory,  it is required that,  
all the  fields  in  double field theory as well as  all of their possible products should be  annihilated by the $\ODD$ d'Alembert operator, $\partial^{2}=\partial_{A}\partial^{A}$.  For an  arbitrary quantity, $\Phi$,  in  double field theory, we require
\be
\partial^{2}\Phi\equiv 0\,.
\label{constraint1}
\ee
Replacing $\Phi$ by $\Phi_{1}\Phi_{2}$, we also have
\be
\partial_{A}\Phi_{1}\partial^{A}\Phi_{2}\equiv 0\,.
\label{constraint2}
\ee
These constraints, which we shall call `the level matching constraints' henceforth,   mean  that the Fourier or momentum  modes form a null space, with respect to the  $\ODD$ metric.  Hence,  the theory is not truly doubled:  there is a choice of coordinates $(x^{\prime},\tx^{\prime})$, related to the original coordinates $(x,\tx)$, by an  
$\ODD$ rotation, in which all the  fields do not depend on the $\tx^{\prime}$ coordinates and the momentum null space consists of the Fourier modes conjugate to $x^{\prime}$ only~\cite{Hohm:2010jy}.  Note that throughout our paper, the equivalence symbol, `$\equiv$', denotes the equality up to the level matching  constraints, (\ref{constraint1}), (\ref{constraint2}).   \\

\noindent  One of the remarkable properties   of the above double field theory action is that, upon the level matching  constraints, it reduces to  (\ref{NSaction}),  the  well-known low energy effective action for  closed string. \\

\noindent Manifestly the double field theory  action as well as  the level matching constraints  are  invariant under the   global  $\ODD$ rotation in a standard manner,
\be
\ba{ll}
\cH_{AB}(y)~\longrightarrow~L_{A}{}^{C}L_{B}{}^{D}\cH_{CD}(y^{\prime})\,,~~~&~~~~d(y)~\longrightarrow~d(y^{\prime})\,,
\ea
\label{ODDglobal}
\ee
where $L\in\ODD$ and $y^{\prime A}$ is the rotated double spacetime coordinates, $y^{\prime A}=y^{B}L_{B}{}^{A}$. The previous T-duality transformation (\ref{TdualityTr}) corresponds to the particular choice, $L=\cT$, mapping the domain of the  theory from the $x$-hyperplane to the $\tx$-hyperplane.

What is less obvious  about  the  double field theory action (\ref{DFTS}) is that it possesses   gauge symmetry, which must be the case  since,  restricted on the $x$-hyperplane, the action (\ref{DFTS}) is nothing but  a rewriting of the effective action (\ref{NSaction}) while  the latter surely  enjoys both the $D$-dimensional diffeomorphism, $x^{\mu}\rightarrow x^{\mu}+\delta x^{\mu}$, and the gauge symmetry of the two form field, $B_{\mu\nu}\rightarrow B_{\mu\nu}+\partial_{[\mu}\Lambda_{\nu]}$.  

 That is to say, while in the effective action (\ref{NSaction}) the gauge symmetry is manifest but  T-duality is not,   in the double field theory action (\ref{DFTS})  it is quite  the opposite.

Nevertheless,  in Ref.\cite{Hohm:2010pp},  Hohm, Hull and Zwiebach showed, through  direct, yet rather lengthy computation\footnote{For another verification, see  a work by Kwak~\cite{Kwak:2010ew}.},  that the double field theory action (\ref{DFTS}) is indeed invariant under the following gauge transformation,
\be
\ba{cll}
\delta_{X}\cH_{AB}&=&X^{C}\partial_{C}\cH_{AB}+(\partial_{A}X_{C}-\partial_{C}X_{A})\cH^{C}{}_{B}+
(\partial_{B}X_{C}-\partial_{C}X_{B})\cH_{A}{}^{C}\,,\\
\delta_{X}d&=&X^{A}\partial_{A}d -\half\partial_{B}X^{B}\,,
\label{cHdTr}
\ea
\ee
where $X^{A}$ is a local gauge parameter, of which half  corresponds to the $D$-dimensional diffeomorphism parameter, $\delta x^{\mu}$, and the other half matches the one-form gauge parameter, $\Lambda_{\nu}$.  The gauge parameter, $X^{A}$,  is also supposed to obey the level matching constraints, (\ref{constraint1}), (\ref{constraint2}), together  with other fields, 
\be
\ba{ll}
\partial^{2}X^{A}\equiv 0\,,~~~~&~~~~\partial_{B}X^{A}\partial^{B}\Phi\equiv 0\,,
\ea
\label{XPhi}
\ee
such that the  constraints are preserved under the gauge transformation, $\partial^{2}\delta_{X}\Phi\equiv 0$.\\

\noindent Especially  the gauge transformation of $\cH_{AB}$ can be identified as   the generalized Lie derivative,\footnote{The transformation of the dilaton, $d$, can be also understood in terms of a modified  generalized Lie derivative designed for `tensor densities',   as explained in section \ref{SECdensity}.}  
\be 
\delta_{X}\cH_{AB}=\hcL_{X}\cH_{AB}\,,
\label{deltacH}
\ee
which  is defined for any field   carrying the $\ODD$ vector indices, such as   scalars, vectors, tensors and ordinary derivatives of them, by~\cite{Grana:2008yw,Hohm:2010pp}
\be
\hcL_{X}T_{A_{1}A_{2}\cdots A_{n}}:=X^{B}\partial_{B}T_{A_{1}A_{2}\cdots A_{n}}+\sum_{i=1}^{n}(\partial_{A_{i}}X^{B}-\partial^{B}X_{A_{i}})T_{A_{1}\cdots A_{i-1}BA_{i+1}\cdots A_{n}}\,.
\label{hcLdef}
\ee
By definition, the \textit{covariant tensors} in double field theory follow the  gauge transformation rule   dictated  by the generalized Lie derivative,
\be
\delta_{X}T_{A_{1}A_{2}\cdots A_{n}}=\hcL_{X}T_{A_{1}A_{2}\cdots A_{n}}\,.
\label{covtensor}
\ee
Since the  constant $\ODD$ invariant metric is annihilated  by the generalized Lie derivative,     $\hcL_{X}\eta_{AB}=0$, in Eq.(\ref{hcLdef}) the lower vector indices, $A_{1}, A_{2},\cdots$, can be freely raised  by the metric, $\eta$. 

The commutator of the  generalized Lie derivatives reads
\be
[\hcL_{X},\hcL_{Y}]=\hcL_{[X,Y]_{\rbC}}+\hat{\cO}_{X,Y}\,,
\label{hcLcom}
\ee
where $[X,Y]_{\rbC}$ denotes the $\mathbf{c}$-bracket introduced by Siegel~\cite{Siegel:1993th},\footnote{Upon the level matching constraints  the  $\mathbf{c}$-bracket reduces to the Courant bracket~\cite{Courant},  as recognized in \cite{Hull:2009zb}.}
\be
[X,Y]^{A}_{\rbC}:=X^{B}\partial_{B}Y^{A}-Y^{B}\partial_{B}X^{A}+\half Y^{B}\partial^{A}X_{B}-\half X^{B}\partial^{A}Y_{B}\,,
\label{Courant}
\ee
%%%
%%\be
%%[X,Y]_{\rbC}:=[X,Y]+\half(Y^{A}\partial^{B}X_{A}-X^{A}\partial^{B}Y_{A})\partial_{B}\,,
%%\label{Courant}
%%\ee
%%%
and $\hat{\cO}_{X,Y}$ is given by
\be
\ba{l}
\hat{\cO}_{X,Y}T_{A_{1}A_{2}\cdots A_{n}}\\
=\half(X^{B}\partial^{C}Y_{B}-Y^{B}\partial^{C}X_{B})\partial_{C}T_{A_{1}A_{2}\cdots A_{n}}
+\dis{\sum_{i=1}^{n}\,}(\partial_{C}Y_{A_{i}}\partial^{C}X^{B}-\partial_{C}X_{A_{i}}\partial^{C}Y^{B})T_{A_{1}\cdots A_{i-1}BA_{i+1}\cdots A_{n}}\,.
\ea
\ee
Thus, imposing the level matching conditions  on the gauge parameters (\ref{XPhi}),  $\hat{\cO}_{X,Y}$ becomes trivial and the gauge algebra is closed by the  $\mathbf{c}$-bracket~\cite{Hull:2009zb},
\be
[\hcL_{X},\hcL_{Y}]\equiv\hcL_{[X,Y]_{\rbC}}\,.
\label{comXYconst}
\ee
It is straightforward to show that the  $\mathbf{c}$-bracket of two covariant vectors is  also an covariant vector, upon the level matching constraints~\cite{Gualtieri:2003dx}, % as $\delta_{X}\!\left([X,Y]^{A}_{\rbC}\right)\equiv\hcL_{X}\!\left([X,Y]^{A}_{\rbC}\right)$. 
\be
\delta_{X}\!\left([X,Y]^{A}_{\rbC}\right)\equiv\hcL_{X}\!\left([X,Y]^{A}_{\rbC}\right)\,.
\label{covCourant}
\ee
~\\

In this paper,  as the underlying mathematical structure of the double field theory, we propose  a  novel differential geometry characterized  by the existence of  a $\ODD$ symmetric projection.  We introduce     a 
 differential operator compatible with the projection, which, contracted  with the projection, can be covariantized and may replace the ordinary derivatives in the generalized Lie derivative (\ref{hcLdef}).  In a systematic fashion, 
we construct  various gauge covariant  tensors. In particular, we reformulate  the double field theory Lagrangian,  $\cR(\cH,d)$ given in (\ref{cRODD}),    in terms of the curvature of our projection-compatible derivative.  Our formalism manifest  both the gauge symmetry and the global $\ODD$ symmetry.\\
~\\
~\\
\textit{Note added}: After submitting the first version  of this manuscript to arXiv, a related interesting  work  by Hohm and Kwak appeared~\cite{Hohm:2010xe}. Their paper also addresses the underlying differential geometry of the double field theory, yet technically differs   from our approach, as they introduce    a covariant derivative whose  connection is not \textit{a priori}  a physical variable of the  double field theory. \\

\newpage

%%%%%%%%%%%%%%%%%%%%%%%%%%%%%%%%%%%%%%%%%%%%%%%%%%%%%%%%%%%%%%%%%%%%%%%%%%%%%%%%%%%%%%%%%%%%%%%%%%%%%%%%%%%%%%%%%%%%%%%%%%%%%%%%%%%%%%%%%%%%%%%%%%%%%%%%%%%%%%%%%%%%%%%%%%
\section{Differential geometry compatible with a projection}
\subsection{Projection}
We start with an observation that the expression of $\cH_{AB}$ in (\ref{gH}) is the most general form of a $2D\times 2D$ matrix 
satisfying
\be
\ba{ll}
\cH^{A}{}_{B}\cH^{B}{}_{C}=\delta^{A}{}_{C}\,,~~~~&~~~~\cH_{AB}=\cH_{BA}\,,
\ea
\ee
and an additional condition that the  upper left $D\times D$ block of $\cH_{AB}$ is non-degenerate. This observation provides a basic  explanation why the  transformation given by $\cH\,\rightarrow\,L\cH L^{t}$, $L\in\ODD$, (\ref{ODDglobal}) leads to the  well-defined gauge transformation of each  component field, $g_{\mu\nu}$ and $B_{\mu\nu}$, as prescribed by Buscher~\cite{Buscher:1985kb,Buscher:1987sk,Buscher:1987qj}.  The second observation to which we pay attention  is that $\cH$ leads to a projection, 
\be
\left[\half(1+\cH)\right]^{A}{}_{B}=\left[\half(1+\cH)\right]^{A}{}_{C}\left[\half(1+\cH)\right]^{C}{}_{B}\,.
\ee

Motivated by the above observations,  henceforth   we focus on  a projection, $P_{A}{}^{B}$, which satisfies  both  the defining relation,\footnote{The positions of the matrix indices should be automatically understood, such that  a matrix multiplication involves a contraction of one upper and one lower index. }
\be
P=P^{2}\,,
\label{prodef}
\ee
and the $\ODD$-symmetric property,
\be
P_{AB}=P_{A}{}^{C}\eta_{CB}=P_{BA}\,.
\label{symmetric}
\ee
We might further demand  that the  upper left $D\times D$ block of ${2P-1}$ is non-degenerate, which would eventually relate the projection to the generalized metric  by   $P=\half(1+\cH)$.   One subtle implication of this technical  assumption of the non-degeneracy would be  $P_{A}{}^{A}=D$, since  Eq.(\ref{gH}) shows that $\cH$ is traceless. The conditions  (\ref{prodef}) and (\ref{symmetric}) alone  do not necessarily  fix the rank of $P$ like that. Nevertheless,  since our main results in the present paper do not care about   any particular value of the rank of the projection,  henceforth  we simply focus on a projection, $P$,  satisfying the two conditions (\ref{prodef}) and (\ref{symmetric}) only.\\

\noindent Conceptually, we place emphasis  on the symmetric \textit{projection}, $P$,  rather than the generalized \textit{metric}, $\cH$. Namely, \textit{in our formalism there is only one metric,}  $\eta$,  that is the constant  $\ODD$ invariant   metric used for raising or lowering the vector indices.\\

\noindent With  the complementary  projection,
\be
\bP:=1-P=\bP^{2}\,,
\ee
 it follows that
\be
\ba{ll}
\partial_{A}P P={\bP}\partial_{A}PP\,,~~~~&~~~~
P\partial_{A}P=P\partial_{A}P{\bP}\,.
\label{pd1}
\ea
\ee
That is to say, if we force to impose a chirality on one free index of $\partial_{A}P$, the other index automatically acquires  the opposite chirality. Therefore, 
\be
\ba{ll}
P\partial_{A}PP=0\,,~~~~&~~~~{\bP}\partial_{A}P{\bP}=0\,.
\ea
\label{pd2}
\ee
Now let us consider a tensor, $\wT_{A_{1}A_{2}\cdots A_{n}}$, of which the $j$th index has a  definite chirality:  With a generic non-chiral tensor carrying vector indices, $T_{A_{1}A_{2}\cdots A_{n}}$, it may be given by
\be
\wT_{A_{1}A_{2}\cdots A_{n}}=(P_{j})_{A_{j}}{}^{B}T_{A_{1}\cdots A_{j-1}BA_{j+1}\cdots A_{n}}\,,
\label{chiralT}
\ee
such that it satisfies   the chirality condition for the $j$th index,
\be
(P_{j})_{A_{j}}{}^{B}\wT_{A_{1}\cdots A_{j-1}BA_{j+1}\cdots A_{n}}=\wT_{A_{1}A_{2}\cdots A_{n}}\,,
\label{jchiral}
\ee
where $P_{j}$ denotes  either $P$ (chiral) or $\bP$ (anti-chiral).  Taking a derivative of  Eq.(\ref{jchiral}),  we get 
\be
\ba{ll}
(P_{j})_{A_{j}}{}^{B}\partial_{C}\wT_{A_{1}\cdots A_{j-1}BA_{j+1}\cdots A_{n}}&=
\partial_{C}\wT_{A_{1}\cdots A_{n}}-(\partial_{C}P_{j})_{A_{j}}{}^{B}\wT_{A_{1}\cdots A_{j-1}BA_{j+1}\cdots A_{n}}\\
{}&=\partial_{C}\wT_{A_{1}\cdots A_{n}}+[\partial_{C}P(\bP-P)P_{j}]_{A_{j}}{}^{B}\wT_{A_{1}\cdots A_{j-1}BA_{j+1}\cdots A_{n}}\\
{}&=\partial_{C}\wT_{A_{1}\cdots A_{n}}+[\partial_{C}P(\bP-P)]_{A_{j}}{}^{B}\wT_{A_{1}\cdots A_{j-1}BA_{j+1}\cdots A_{n}}\,.
\ea
\label{derjc}
\ee
Motivated by  this  simple exercise,\footnote{We may also consider  a tensor, $\wT_{A_{1}A_{2}\cdots A_{n}}$, of which every index has a definite chirality such that 
\[
\wT_{A_{1}A_{2}\cdots A_{n}}=(P_{1})_{A_{1}}{}^{B_{1}}(P_{2})_{A_{2}}{}^{B_{2}}\cdots (P_{n})_{A_{n}}{}^{B_{n}}\wT_{B_{1}B_{2}\cdots B_{n}}\,,
\]
where for each $i$th component ($i=1,2,\cdots,n$), $P_{i}$ is either $P$ (chiral) or $\bP$ (anti-chiral).  Taking a derivative of this gives an expression,
\[
\partial_{C}\wT_{A_{1}A_{2}\cdots A_{n}}-
\sum_{i=1}^{n}(\partial_{C}P_{i})_{A_{i}}{}^{B}\wT_{A_{1}\cdots A_{i-1}BA_{i+1}\cdots A_{n}}
=(P_{1})_{A_{1}}{}^{B_{1}}(P_{2})_{A_{2}}{}^{B_{2}}\cdots (P_{n})_{A_{n}}{}^{B_{n}}\partial_{C}\wT_{B_{1}B_{2}\cdots B_{n}}\,,
\]
 which also suggests the form of  $\nD_{C}$ defined  in (\ref{nDdef}). }  as a preliminary result,   we may define a non-covariant `chirality preserving derivative', $\nD_{C}$, which acts  on a  generic (not necessarily chiral) tensor, $T_{A_{1}A_{2}\cdots A_{n}}$, 
%%%
%%\be
%%\nD_{C}T_{A_{1}A_{2}\cdots A_{n}}:=\partial_{C}T_{A_{1}A_{2}\cdots A_{n}}-
%%\sum_{i=1}^{n}\,[\partial_{C}P(P-\bP)]_{A_{i}}{}^{B}T_{A_{1}\cdots A_{i-1}BA_{i+1}\cdots A_{n}}\,.
%%\label{nDdef}
%%\ee
%%%
\be
\nD_{C}T_{A_{1}A_{2}\cdots A_{n}}:=\partial_{C}T_{A_{1}A_{2}\cdots A_{n}}+
\sum_{i=1}^{n}\,\nDGamma_{CA_{i}}{}^{B}\,T_{A_{1}\cdots A_{i-1}BA_{i+1}\cdots A_{n}}\,,
\label{nDdef}
\ee
where we set the connection to be%\footnote{In our convention,  the anti-symmetrization of the superscript or subscript  indices is taken with the weight one, such that $T_{[AB]}=\half(T_{AB}-T_{BA})$, \textit{etc.}}
\be
\nDGamma_{CAB}:=[\partial_{C}P(\bP-P)]_{AB}=-[\partial_{C}P(\bP-P)]_{BA}=-2(\partial_{C}PP)_{[AB]}\,.
\label{nDGdef}
\ee
In our convention,  the symmetrization and the anti-symmetrization of the superscript or subscript  indices are taken with the weight one, such that $T_{(AB)}:=\half(T_{AB}+T_{BA})$, $\,T_{[AB]}:=\half(T_{AB}-T_{BA})$, \textit{etc.}\\

\noindent With  (\ref{pd1}) one can check easily,
\be
\ba{ll}
\nD_{A}\eta_{BC}=0\,,~~~~&~~~~\nD_{A}P_{BC}=0\,.
\ea
\label{pseudoconstant}
\ee
Hence, the derivative preserves any existing  chirality: for  the chiral tensor, $\wT_{A_{1}\cdots A_{n}}$, given in (\ref{chiralT}), we get
\be
\nD_{C}\wT_{A_{1}A_{2}\cdots A_{n}}=(P_{j})_{A_{j}}{}^{B}\nD_{C}\wT_{A_{1}\cdots A_{j-1}BA_{j+1}\cdots B_{n}}\,.
\label{nDchiral}
\ee
However, under the gauge   transformation (\ref{covtensor}), this derivative  transforms non-covariantly: Straightforward  computation  shows, for an arbitrary generic tensor, $T_{A_{1}\cdots A_{n}}$, 
\be
\ba{ll}
(\delta_{X}-\hcL_{X})(\nD_{C}T_{A_{1}\cdots A_{n}})=&\sum_{i}\, {\bP}_{A_{i}}{}^{D}(\partial_{C}\partial_{D}X^{E}-\partial_{C}\partial^{E}X_{D}){\bP}_{E}{}^{B}T_{A_{1}\cdots A_{i-1}BA_{i+1}\cdots A_{n}}\\
{}&+\sum_{i}\,P_{A_{i}}{}^{D}(\partial_{C}\partial_{D}X^{E}-\partial_{C}\partial^{E}X_{D})P_{E}{}^{B}T_{A_{1}\cdots A_{i-1}BA_{i+1}\cdots A_{n}}\\
{}&+\partial^{B}X_{C}\nD_{B}T_{A_{1}\cdots A_{n}}\,,
\ea
\label{noncovprime}
\ee
where the last term involving $\partial^{B}X_{C}\nD_{B}$ vanishes thanks to the level matching  constraint (\ref{constraint2}), yet the others survive as nontrivial   inhomogeneous terms. In order to cancel them, we need to add  extra terms to  the connection of $\nD_{c}$, \textit{i.e.~}$\nDGamma_{CAB}$ in (\ref{nDGdef}).\footnote{ In fact, the most general form of the connection satisfying the two conditions in Eq.(\ref{pseudoconstant}) is given by  
\[\nDGamma_{CAB}+\Omega_{C[AB]}+(P-\bP)_{A}{}^{D}(P-\bP)_{B}{}^{E}\Omega_{C[DE]}\,,\] where $\Omega_{CAB}$ is arbitrary.} The  new connection we look for  should be constructed from the projection,  and preserve any existing  chiral structure of a tensor. Due to (\ref{pd2}), there are essentially two candidate pieces which we may   add to  the connection  in (\ref{nDGdef}): for the chiral index,
\be
P_{D[A}(P\partial^{D}P)_{B]C}=P_{D[A}[P\partial^{D}P{\bP}]_{B]C}\,,
\label{chiralcandi}
\ee
and for the anti-chiral index,
\be
{\bP}_{D[A}(\partial^{D}PP)_{B]C}={\bP}_{D[A}[{\bP}\partial^{D}PP)]_{B]C}\,.
\label{antichiralcandi}
\ee
Note that because of (\ref{pd1}), imposing the same chiralities on $A$ and $B$ indices forces the remaining index, $C$, to assume  the opposite chirality. \\~\\
%%%
%%under the  local   transformation as
%%\be
%%(\delta_{X}-\hcL_{X})P_{D[A}(P\partial^{D}P)_{B]C}=
%%\partial_{D}X^{E}P_{E[A}(P\partial^{D}P)_{B]C}+
%% P_{A}{}^{D}P_{B}{}^{E}{\bP}_{C}{}^{F}\partial_{F}\partial_{[D}X_{E]}\,,
%% \ee
%% and
%% \be
%% (\delta_{X}-\hcL_{X}){\bP}_{D[A}(\partial^{D}PP)_{B]C} =
%%\partial_{D}X^{E}{\bP}_{E[A}(\partial^{D}PP)_{B]C}-
%% {\bP}_{A}{}^{D}{\bP}_{B}{}^{E}P_{C}{}^{F}\partial_{F}\partial_{[D}X_{E]}\,.
%%\ee
%%%
%%%%
%%%%%%%%%%%%%%%%%%%%%%%%%%%%%%%%%%%%%%%%%%%%%%%%%%%%%%%%%%%%%%%%%%%%%%%%%%%%%%%%%%%%%%%%%%%%%%%%%%%%%%%%%%%%%%%%%%%%%%%%%%%%%%%%%%%%%%%%%%%%%%%%%%%%%%%%%%%%%%%%%%%%%%%%%%
\subsection{Projection-compatible derivative}
Based on the  previous preliminary analysis,   we  define the following    \textit{projection-compatible derivative},  $D_{C}$, which acts  on a  generic field carrying $\ODD$ vector indices (not necessarily a covariant tensor) as
\be
D_{C}T_{A_{1}A_{2}\cdots A_{n}}:=\partial_{C}T_{A_{1}A_{2}\cdots A_{n}}+
\sum_{i=1}^{n}\,\Gamma_{CA_{i}}{}^{B}T_{A_{1}\cdots A_{i-1}BA_{i+1}\cdots A_{n}}\,,
\label{Ddef}
\ee
where the connection is 
\be
\ba{ll}
\Gamma_{CAB}&=[\partial_{C}P(\bP-P)]_{AB}-2P_{D[A}(P\partial^{D}P)_{B]C}+2{\bP}_{D[A}(\partial^{D}PP)_{B]C}\\
{}&=[\partial_{C}P(1-2P)]_{AB}-P_{A}{}^{D}\partial_{D}P_{BC}+(\partial_{A}PP)_{BC}+P_{B}{}^{D}\partial_{D}P_{AC}-(\partial_{B}PP)_{AC}\\
{}&=2P_{[A}{}^{D}{\bP}_{B]}{}^{E}\partial_{C}P_{DE}+2\left({\bP}_{[A}{}^{D}{\bP}_{B]}{}^{E}-P_{[A}{}^{D}P_{B]}{}^{E}\right)\partial_{D}P_{EC}\,.
\ea
\label{connectionG}
\ee
This  connection  is a  unique combination of  the terms,  (\ref{nDGdef}), (\ref{chiralcandi}) and (\ref{antichiralcandi}), to satisfy
\be
\ba{ll}
\Gamma_{CAB}+\Gamma_{CBA}=0\,,~~~~&~~~~
\Gamma_{ABC}+\Gamma_{CAB}+\Gamma_{BCA}=0\,.
\ea
\label{sympropG}
\ee
Thanks to these two symmetric properties,  all the ordinary derivatives  in the definitions  of the generalized Lie derivative (\ref{hcLdef}) as well as  the $\mathbf{c}$-bracket (\ref{Courant}) can be now  replaced by our projection-compatible  derivatives:
\be
\ba{ll}
\hcL_{X}T_{A_{1}A_{2}\cdots A_{n}}&=X^{B}\partial_{B}T_{A_{1}A_{2}\cdots A_{n}}+\sum_{i=1}^{n}(\partial_{A_{i}}X_{B}-\partial_{B}X_{A_{i}})T_{A_{1}\cdots A_{i-1}}{}^{B}{}_{A_{i+1}\cdots A_{n}}\\
{}&=X^{B}D_{B}T_{A_{1}A_{2}\cdots A_{n}}+\sum_{i=1}^{n}(D_{A_{i}}X_{B}-D_{B}X_{A_{i}})T_{A_{1}\cdots A_{i-1}}{}^{B}{}_{A_{i+1}\cdots A_{n}}\,,
\ea
\label{ordiproD}
\ee
and
\be
\ba{ll}
[X,Y]^{A}_{\rbC}&=X^{B}\partial_{B}Y^{A}-Y^{B}\partial_{B}X^{A}+\half Y^{B}\partial^{A}X_{B}-\half X^{B}\partial^{A}Y_{B}\\
{}&=X^{B}D_{B}Y^{A}-Y^{B}D_{B}X^{A}+\half Y^{B}D^{A}X_{B}-\half X^{B}D^{A}Y_{B}\,.
\ea
\label{CourantD}
\ee

Like (\ref{pseudoconstant}), both the $\ODD$ invariant constant metric and the symmetric projection are `constant' with respect to the derivative:\footnote{Since the connection is $\sodd$ valued as $\Gamma_{CAB}=-\Gamma_{CBA}$, the  Levi-Civita symbol, $\epsilon^{A_{1}A_{2}\cdots A_{2D}}$,  is also constant with respect to the derivative, 
$D_{B}\epsilon^{A_{1}A_{2}\cdots A_{2D}}=0$.}
\be
\ba{ll}
D_{A}\eta_{BC}=0\,,~~~~&~~~~D_{A}P_{BC}=0\,.
\ea
\label{pseudoconstant2}
\ee
The derivative, $D_{A}$, is compatible with the projection, such that,  like (\ref{nDchiral}), the derivative preserves any existing chirality: As in (\ref{chiralT})  for 
\be
\wT_{A_{1}A_{2}\cdots A_{n}}=(P_{j})_{A_{j}}{}^{B}\wT_{A_{1}\cdots A_{j-1}BA_{j+1}\cdots A_{n}}\,,
\ee
we note
\be
D_{C}\wT_{A_{1}A_{2}\cdots A_{n}}=(P_{j})_{A_{j}}{}^{B}D_{C}\wT_{A_{1}\cdots A_{j-1}BA_{j+1}\cdots B_{n}}\,.
\label{Dchiral}
\ee
~\\

Under an arbitrary infinitesimal transformation of the projection satisfying from (\ref{pd2})
\be
\delta P= P\delta P\bP+\bP\delta P P\,,
\label{infPtr}
\ee
the connection transforms as
\be
\ba{ll}
\delta\Gamma_{CAB}=&2P_{[A}{}^{D}\bP_{B]}{}^{E}D_{C}\delta P_{DE}+2(\bP_{[A}{}^{D}\bP_{B]}{}^{E}-
P_{[A}{}^{D}P_{B]}{}^{E})D_{D}\delta P_{EC}\\
{}&{}-\Gamma_{FDE\,}\delta(P_{C}{}^{F}P_{A}{}^{D}P_{B}{}^{E}+
\bP_{C}{}^{F}\bP_{A}{}^{D}\bP_{B}{}^{E})\,.
\ea
\label{infGtr}
\ee
Especially under the  gauge  transformation (\ref{deltacH}), (\ref{covtensor}),  the connection transforms as
\be
(\delta_{X}-\hcL_{X})\Gamma_{CAB}
\equiv \left(P_{A}{}^{D}P_{B}{}^{E}P_{C}{}^{F}+
{\bP}_{A}{}^{D}{\bP}_{B}{}^{E}{\bP}_{C}{}^{F}-\delta_{A}{}^{D}\delta_{B}{}^{E}\delta_{C}{}^{F}\right)(\partial_{F}\partial_{D}X_{E}-\partial_{F}\partial_{E}X_{D})\,,
\label{Godd}
\ee
such that our projection-compatible derivative transforms as
\be
\left(\delta_{X}-\hcL_{X}\right)D_{C}T_{A_{1}A_{2}\cdots A_{n}}\equiv
2\sum_{i=1}^{n}\left(P_{A_{i}}{}^{D}P_{B}{}^{E}P_{C}{}^{F}+{\bP}_{A_{i}}{}^{D}{\bP}_{B}{}^{E}{\bP}_{C}{}^{F}\right)
\partial_{F}\partial_{[D}X_{E]}T_{A_{1}\cdots A_{i-1}}{}^{B}{}_{A_{i+1}\cdots A_{n}}\,.
\label{DTodd}
\ee
In comparison with   (\ref{noncovprime}), the main difference here is that  all the vector indices of the lefthand side of (\ref{DTodd}) appear through the projections on the righthand side.  Thus,  
 the following two quantities are gauge covariant  tensors, 
\be
\ba{l}
P_{C}{}^{D}{\bP}_{A_{1}}{}^{B_{1}}{\bP}_{A_{2}}{}^{B_{2}}\cdots{\bP}_{A_{n}}{}^{B_{n}}
D_{D}T_{B_{1}B_{2}\cdots B_{n}}\,,\\
{\bP}_{C}{}^{D}P_{A_{1}}{}^{B_{1}}P_{A_{2}}{}^{B_{2}}\cdots P_{A_{n}}{}^{B_{n}}
D_{D}T_{B_{1}B_{2}\cdots B_{n}}\,.
\ea
\label{PDPT}
\ee
Namely, combined with the projections,  our projection-compatible  derivative, $D_{A}$ in  (\ref{Ddef}), gives rise to covariant derivatives. \\

\noindent For later use, it is worth while  to note that the successive use of (\ref{DTodd}) gives
\be
\ba{l}
(\delta_{X}-\hcL_{X})D_{A}D_{B}T_{C_{1}C_{2}\cdots C_{n}}\\
\equiv ~2\left(P_{A}{}^{D}P_{B}{}^{E}P_{F}{}^{G}+
{\bP}_{A}{}^{D}{\bP}_{B}{}^{E}{\bP}_{F}{}^{G}\right)\partial_{D}\partial_{[E}X_{G]}D^{F}T_{C_{1}C_{2}\cdots C_{n}}\\
{}~~~~+\sum_{i}2\left(P_{C_{i}}{}^{D}P_{F}{}^{G}P_{B}{}^{E}+
{\bP}_{C_{i}}{}^{D}{\bP}_{F}{}^{G}{\bP}_{B}{}^{E}\right)(D_{A}\partial_{E}\partial_{[D}X_{G]})T_{C_{1}\cdots C_{i-1}}{}^{F}{}_{C_{i+1}\cdots C_{n}}\\
{}~~~~+\sum_{i}2\left(P_{C_{i}}{}^{D}P_{F}{}^{G}P_{B}{}^{E}+
{\bP}_{C_{i}}{}^{D}{\bP}_{F}{}^{G}{\bP}_{B}{}^{E}\right)
\partial_{E}\partial_{[D}X_{G]}\,D_{A}T_{C_{1}\cdots C_{i-1}}{}^{F}{}_{C_{i+1}\cdots C_{n}}\\
{}~~~~+\sum_{i}2\left(P_{C_{i}}{}^{D}P_{F}{}^{G}P_{A}{}^{E}+
{\bP}_{C_{i}}{}^{D}{\bP}_{F}{}^{G}{\bP}_{A}{}^{E}\right)
\partial_{E}\partial_{[D}X_{G]}\,D_{B}T_{C_{1}\cdots C_{i-1}}{}^{F}{}_{C_{i+1}\cdots C_{n}}\,.
\ea
\label{DDT}
\ee
~\\

%%%%%%%%%%%%%%%%%%%%%%%%%%%%%%%%%%%%%%%%%%%%%%%%%%%%%%%%%%%%%%%%%%%%%%%%%%%%%%%%%%%%%%%%%%%%%%%%%%%%%%%%%%%%%%%%%%%%%%%%%%%%%%%%%%%%%%%%%%%%%%%%%%%%%%%%%%%%%%%%%%%%%%%%%%
\subsection{Curvature}
The commutator of the  projection-compatible  derivatives (\ref{Ddef}) reads
\be
{}[D_{A},D_{B}]T_{C_{1}C_{2}\cdots C_{n}}=-\Gamma_{DAB}D^{D}T_{C_{1}C_{2}\cdots C_{n}}+
\sum_{i=1}^{n}R_{C_{i}DAB}\,T_{C_{1}\cdots C_{i-1}}{}^{D}{}_{C_{i+1}\cdots C_{n}}\,,
\label{Rcomm}
\ee
where, from (\ref{sympropG}),  $\Gamma_{DAB}=\Gamma_{ABD}-\Gamma_{BAD}$ corresponds to the torsion and  $R_{CDAB}$ is the curvature given by, following the standard convention in Riemannian geometry,  
\be
\ba{lrl}
R_{CDAB}&:=&\partial_{A}\Gamma_{BCD}-\partial_{B}\Gamma_{ACD}+\Gamma_{AC}{}^{E}\Gamma_{BED}-\Gamma_{BC}{}^{E}\Gamma_{AED}\\
{}&=&D_{A}\Gamma_{BCD}-D_{B}\Gamma_{ACD}+\Gamma_{EAB}\Gamma^{E}{}_{CD}+\Gamma_{ACE}\Gamma_{BD}{}^{E}-\Gamma_{ADE}\Gamma_{BC}{}^{E}\,.
\ea
\ee
This curvature is anti-symmetric for the first two and also for the last two indices respectively,
\be
R_{CDAB}=R_{[CD][AB]}\,.
\ee
From (\ref{pseudoconstant2}) and (\ref{Rcomm}), it follows that
\be
P_{C}{}^{E}R_{EDAB}=P_{D}{}^{F}R_{CFAB}\,.
\ee
In particular,
\be
P_{C}{}^{E}{\bP}_{D}{}^{F}R_{EFAB}=0\,.
\ee
The  Jacobi identity,
\be
[D_{A},[D_{B},D_{C}]]+[D_{C},[D_{A},D_{B}]]+[D_{B},[D_{C},D_{A}]]=0\,,
\ee
yields 
\be
R_{[A}{}^{D}{}_{BC]}+D_{[A}\Gamma^{D}{}_{BC]}+\Gamma^{D}{}_{E[A}\Gamma^{E}{}_{BC]}=0\,,
\label{asR}
\ee
and
\be
D_{[A}R^{DE}{}_{BC]}+R^{DE}{}_{F[A}\Gamma^{F}{}_{BC]}=0\,.
\ee
Contracting some of the vector indices,    the latter leads to an identity,
\be
D_{A}(R^{AB}-\half\eta^{AB}R)=\half R^{CDAB}\Gamma_{ACD}-R_{CD}\Gamma^{DCB}\,,
\ee
where $R_{AB}-\half\eta_{AB}R$ is reminiscent of  the familiar  Einstein tensor.  Note that throughout our paper we set, following the standard convention, 
\be
\ba{ll}
R_{AB}:=R^{C}{}_{ACB}\,,~~~~&~~~~R:=R^{A}{}_{A}=R^{AB}{}_{AB}\,.
\ea
\ee
Unlike the Ricci curvature  in the ordinary Riemann geometry, our $R_{AB}$ is not symmetric, 
\be
R_{AB}\neq R_{BA}\,.
\ee

Further,  we define  as for a key quantity  in  the presentation of our main results later ( \textit{cf.}~\cite{Siegel:1993th} ),
\be
S_{ABCD}:=\half\left(R_{ABCD}+R_{CDAB}-\Gamma^{E}{}_{AB}\Gamma_{ECD}\right)\,,
\ee
which satisfies,  with (\ref{asR}),  all the symmetric properties of the standard Riemann curvature, 
\be
\ba{lll}
S_{ABCD}=S_{[AB][CD]}\,,~~~~&~~~~S_{ABCD}\equiv S_{CDAB}\,,~~~~&~~~~S_{A[BCD]}=0\,,
\ea
\label{symS}
\ee
as well as,  from  brute force computation,
\be
\ba{ll}
P_{A}{}^{E}\bP_{B}{}^{F}P_{C}{}^{G}\bP_{D}{}^{H}S_{EFGH}\equiv 0\,,~~~~&~~~~P_{A}{}^{E}P_{B}{}^{F}\bP_{C}{}^{G}\bP_{D}{}^{H}S_{EFGH}\equiv 0\,.
\ea
\label{ppppS}
\ee
In fact, if one computes the commutator of  the  generalized Lie derivatives,  in terms of the projection-compatible  derivatives (\ref{ordiproD}), one obtains
\be
\left([\hcL_{X},\hcL_{Y}]-\hcL_{[X,Y]_{\rbC}}-\hat{\cO}_{X,Y}\right)T_{A_{1}A_{2}\cdots A_{n}}=\dis{\sum_{i=1}^{n}\,}
6S_{A_{i}[BCD]}X^{B}Y^{C}T_{A_{1}\cdots A_{i-1}}{}^{D}{}_{A_{i+1}\cdots A_{n}}\,,
\ee
of which the right hand side vanishes, due to the latter identity of (\ref{symS}), such that the result is consistent with (\ref{hcLcom}).  It follows from  (\ref{ppppS})  that, if we set
\be
F_{ABCD}:=\left(P_{A}{}^{E}P_{B}{}^{F}\bP_{C}{}^{G}\bP_{D}{}^{H}-\bP_{A}{}^{E}\bP_{B}{}^{F}P_{C}{}^{G}P_{D}{}^{H}\right)R_{EFGH}\,,
\ee
then $F_{ABCD}$ also satisfies the two symmetric properties,
\be
\ba{ll}
F_{ABCD}=F_{[AB][CD]}\,,~~~~&~~~~F_{ABCD}\equiv F_{CDAB}\,.
\ea
\ee
~\\

Either through direct  calculation  or alternatively by considering the gauge transformation of   (\ref{Rcomm}) with (\ref{DDT}),  one can obtain the gauge transformation of   the curvature,
\be
\ba{ll}
(\delta_{X}-\hcL_{X})R_{CDAB}\equiv& -2\Gamma^{E}{}_{CD}\partial_{E}\partial_{[A}X_{B]}\\
{}&+P_{C}{}^{E}P_{D}{}^{F}
\left(P^{G}{}_{H}\Gamma^{H}{}_{AB}+P^{G}{}_{B}D_{A}-P^{G}{}_{A}D_{B}\right)
2\partial_{G}\partial_{[E}X_{F]}\\
{}&+{\bP}_{C}{}^{E}{\bP}_{D}{}^{F}
\left({\bP}^{G}{}_{H}\Gamma^{H}{}_{AB}+{\bP}^{G}{}_{B}D_{A}-{\bP}^{G}{}_{A}D_{B}\right)2\partial_{G}\partial_{[E}X_{F]}\,.
\ea
\ee
This result implies 
\be
\ba{l}
(\delta_{X}-\hcL_{X})S_{ABCD}\\
%\!\!=(\delta_{X}-\hcL_{X})\left(R_{ABCD}+R_{CDAB}-\Gamma^{E}{}_{AB}\Gamma_{ECD}\right)\\
\!\!\equiv-\left(P_{A}{}^{E}P_{B}{}^{F}P_{[C}{}^{G}D_{D]}\!+P_{C}{}^{E}P_{D}{}^{F}P_{[A}{}^{G}D_{B]}\!+\bP_{A}{}^{E}\bP_{B}{}^{F}\bP_{[C}{}^{G}D_{D]}\!+\bP_{C}{}^{E}\bP_{D}{}^{F}\bP_{[A}{}^{G}D_{B]}
\right)2\partial_{G}\partial_{[E}X_{F]}\,,
\ea
\ee
such that, contracting some indices as
\be
S_{AB}:=S^{C}{}_{ACB}=S_{BA}=R_{(AB)}-\half\Gamma^{CD}{}_{A}\Gamma_{CDB}\,,
\ee
and
\be
{}~S:=S^{A}{}_{A}=S^{AB}{}_{AB}=R-\half\Gamma^{ABC}\Gamma_{ABC}\,,~~~~~~~~~~~~
\ee
we have
\be
\ba{l}
(\delta_{X}-\hcL_{X})S_{AB}\\
\equiv \left(P_{(A}{}^{C}P_{B)}{}^{D}P^{EF}D_{F}-P^{CE}P_{(A}{}^{D}D_{B)}+
\bP_{(A}{}^{C}\bP_{B)}{}^{D}\bP^{EF}D_{F}-\bP^{CE}\bP_{(A}{}^{D}D_{B)}
\right)2\partial_{C}\partial_{[E}X_{D]}\,,
\ea
\label{RABsym}
\ee
and
\be
(\delta_{X}-\hcL_{X})S\equiv 4\left(P^{CD}P^{AB}+\bP^{CD}\bP^{AB}\right)D_{A}\partial_{C}\partial_{[B}X_{D]}\,.
\ee
~\\

In general, under any infinitesimal transformation of the connection, which may be given by (\ref{infGtr}),  $S_{ABCD}$ transforms as
\be
\delta S_{ABCD}=D_{[A}\delta\Gamma_{B]CD}+D_{[C}\delta\Gamma_{D]AB}\,.
\label{infStr}
\ee
~\\

%%%%%%%%%%%%%%%%%%%%%%%%%%%%%%%%%%%%%%%%%%%%%%%%%%%%%%%%%%%%%%%%%%%%%%%%%%%%%%%%%%%%%%%%%%%%%%%%%%%%%%%%%%%%%%%%%%%%%%%%%%%%%%%%%%%%%%%%%%%%%%%%%%%%%%%%%%%%%%%%%%%%%%%%%%
\subsection{Generalization to tensor densities and dilaton\label{SECdensity}}
In this subsection, we consider the following modification of the generalized Lie derivative,
\be
\ba{lrl}
\tcL_{X}\Tw_{A_{1}A_{2}\cdots A_{n}}\!\!&:=&\!\!\hcL_{X}\Tw_{A_{1}A_{2}\cdots A_{n}}+\omega\partial_{B}X^{B}\Tw_{A_{1}A_{2}\cdots A_{n}}\\
{}\!\!&=&\!\! X^{B}\partial_{B}\Tw_{A_{1}A_{2}\cdots A_{n}}+\omega\partial_{B}X^{B}\Tw_{A_{1}A_{2}\cdots A_{n}} +\dis{\sum_{i=1}^{n}}\,2\partial_{[A_{i}}X_{B]}\Tw_{A_{1}\cdots A_{i-1}}{}^{B}{}_{A_{i+1}\cdots A_{n}}\,,
\ea
\label{hcLdefwT}
\ee
where $\omega$ is the weight of each field, $\Tw_{A_{1}A_{2}\cdots A_{n}}$.  Letting this be the gauge transformation of the field, 
\be
\delta_{X}\Tw_{A_{1}A_{2}\cdots A_{n}}=\tcL_{X}\Tw_{A_{1}A_{2}\cdots A_{n}}\,,
\ee
the field,  $\Tw_{A_{1}A_{2}\cdots A_{n}}$ is identified as a \textit{tensor density}  with weight, $\omega$. The aforementioned   covariant tensors then have the weight zero.\footnote{As the covariant tensors have the trivial weight, one might wish to unite $\tcL_{X}$ with $\hcL_{X}$, and use a single symbol for the Lie derivative. However, it appears that keeping them separately provides less confusing notation, especially for the higher order derivatives of the  dilaton. For example, in our convention,
\[D_{A}\nabla_{B}d=\partial_{A}\nabla_{B}d+\Gamma_{AB}{}^{C}\nabla_{C}d\,.
\] } \\

\noindent Like (\ref{comXYconst}), up to the level matching constraints,  the commutator of them is still closed by  the $\mathbf{c}$-bracket   (\ref{Courant}), 
\be
\ba{ll}
{}[\tcL_{X},\tcL_{Y}]\equiv\tcL_{[X,Y]_{\rbC}}\,,~~&~~
{}[X,Y]^{A}_{\rbC}=X^{B}\partial_{B}Y^{A}-Y^{B}\partial_{B}X^{A}+\half Y^{B}\partial^{A}X_{B}-\half X^{B}\partial^{A}Y_{B}\,.
%%%
%%[X,Y]_{\rbC}=[X,Y]+\half(Y^{A}\partial^{B}X_{A}-X^{A}\partial^{B}Y_{A})\partial_{B}\,.
%%%
\ea
\ee
Furthermore, with the  projection-compatible derivative, $D_{C}$ in (\ref{Ddef}), if we set
\be
\ba{lrl}
\nabla_{C}\Tw_{A_{1}A_{2}\cdots A_{n}}&:=&
D_{C}\Tw_{A_{1}A_{2}\cdots A_{n}}-\omega\Gamma^{B}{}_{BC}\Tw_{A_{1}A_{2}\cdots A_{n}}\\
{}&=&\partial_{C}\Tw_{A_{1}A_{2}\cdots A_{n}}-\omega\Gamma^{B}{}_{BC}\Tw_{A_{1}A_{2}\cdots A_{n}}+
\sum_{i=1}^{n}\,\Gamma_{CA_{i}}{}^{B}\Tw_{A_{1}\cdots A_{i-1}BA_{i+1}\cdots A_{n}}\,,
\ea
\label{DdefwT}
\ee
like (\ref{ordiproD}),  all the ordinary derivatives  in the definition of the modified generalized Lie derivative (\ref{hcLdefwT}) can be replaced by $\nabla_{C}$,
\be
\ba{lll}
\tcL_{X}\Tw_{A_{1}\cdots A_{n}}&=&X^{B}\partial_{B}\Tw_{A_{1}\cdots A_{n}}~+\omega\partial_{B}X^{B}\Tw_{A_{1}\cdots A_{n}} ~+\sum_{i=1}^{n}2\partial_{[A_{i}}X_{B]}\Tw_{A_{1}\cdots A_{i-1}}{}^{B}{}_{A_{i+1}\cdots A_{n}}\\
{}&=&X^{B}\nabla_{B}\Tw_{A_{1}\cdots A_{n}}+\omega\nabla_{B}X^{B}\Tw_{A_{1}\cdots A_{n}} +\sum_{i=1}^{n}2\nabla_{[A_{i}}X_{B]}\Tw_{A_{1}\cdots A_{i-1}}{}^{B}{}_{A_{i+1}\cdots A_{n}}\,.
\ea
\label{ordinabla}
\ee
Since the dilaton, $d$, is the logarithm of a scalar density with weight one (\ref{dphi}),
\be
d=-\half\ln\left(\sqrt{-g}e^{-2\phi}\right)\,,
\ee
from the consideration, 
\be
\nabla_{A}e^{-2d}=(-2\nabla_{A}d)e^{-2d}\,,
\ee
the definition of $\nabla_{A}d\,$  follows naturally,
\be
\nabla_{A}d:=\partial_{A}d+\half\Gamma^{B}{}_{BA}\,.
\label{nablad}
\ee
Like (\ref{ordinabla}),  all the ordinary derivatives  in the gauge transformation of the dilaton given in 
(\ref{cHdTr})  can be replaced by $\nabla_{C}$,
\be
\delta_{X}d=X^{A}\partial_{A}d -\half\partial_{B}X^{B}=X^{A}\nabla_{A}d -\half\nabla_{B}X^{B}\,.
\label{deltadordinabla}
\ee
~\\

For later use, we note 
\be
(\delta_{X}-\hcL_{X})\nabla_{A}d\equiv\left(P_{A}{}^{B}P^{CD}+{\bP}_{A}{}^{B}{\bP}^{CD}\right)\partial_{C}\partial_{[D}X_{B]}\,,
\ee
and
\be
\ba{ll}
(\delta_{X}-\hcL_{X})D_{A}\nabla_{B}d&=(\delta_{X}-\hcL_{X})(\partial_{A}\nabla_{B}d+\Gamma_{AB}{}^{C}\nabla_{C}d)\\
{}&\equiv
2\left(P_{A}{}^{C}P_{B}{}^{D}P^{EF}+\bP_{A}{}^{C}\bP_{B}{}^{D}\bP^{EF}\right)(\nabla_{E}d)\, \partial_{C}\partial_{[D}X_{F]}\\
{}&~~~~+
\left(P_{B}{}^{D}P^{EF}+\bP_{B}{}^{D}\bP^{EF}\right)D_{A}\partial_{E}\partial_{[F}X_{D]}\,.
\ea
\label{Dnablad}
\ee
~\\
~\\
~\\

%%%%%%%%%%%%%%%%%%%%%%%%%%%%%%%%%%%%%%%%%%%%%%%%%%%%%%%%%%%%%%%%%%%%%%%%%%%%%%%%%%%%%%%%%%%%%%%%%%%%%%%%%%%%%%%%%%%%%%%%%%%%%%%%%%%%%%%%%%%%%%%%%%%%%%%%%%%%%%%%%%%%%%%%%%
\section{Gauge covariant tensors: \textit{summary}\label{SECsummary}}
In summary, assembling all the  jigsaw puzzle pieces obtained above,  the following quantities are  gauge covariant,  $\ODD$ tensors, in addition to the $\mathbf{c}$-bracket  (\ref{covCourant}).

\begin{itemize}
%%%
%%\item  Recalling (\ref{covCourant}), the $\mathbf{c}$-bracket of two covariant vectors,
%%\be
%%[X,Y]^{A}_{\rbC}=X^{B}D_{B}Y^{A}-Y^{B}D_{B}X^{A}+\half Y^{B}D^{A}X_{B}-\half X^{B}D^{A}Y_{B}\,.
%%\ee
%%%

\item For a generic gauge covariant tensor, $T_{A_{1}A_{2}\cdots A_{n}}$, recalling Eq.(\ref{PDPT}),
\begin{eqnarray}
&&P_{C}{}^{D}{\bP}_{A_{1}}{}^{B_{1}}{\bP}_{A_{2}}{}^{B_{2}}\cdots{\bP}_{A_{n}}{}^{B_{n}}
D_{D}T_{B_{1}B_{2}\cdots B_{n}}\,,\\
&&{\bP}_{C}{}^{D}P_{A_{1}}{}^{B_{1}}P_{A_{2}}{}^{B_{2}}\cdots P_{A_{n}}{}^{B_{n}}
D_{D}T_{B_{1}B_{2}\cdots B_{n}}\,.
\end{eqnarray}

\item  For a   gauge covariant vector, $V_{A}$,
\begin{eqnarray}
&&P^{AB}(D_{A}-2\nabla_{A}d)V_{B}\,,\\
&&\bP^{AB}(D_{A}-2\nabla_{A}d)V_{B}\,.
\end{eqnarray}
Or equivalently, with $\cH_{AB}=P_{AB}-\bP_{AB}$,
\begin{eqnarray}
&(D_{A}-2\nabla_{A}d)V^{A}\,,\\
&\cH^{AB}(D_{A}-2\nabla_{A}d)V_{B}\,,
\end{eqnarray}
which may be viewed as \textit{gauge covariant divergences} of the vector.

\item Second order derivatives:
\begin{eqnarray}
&&P^{AB}{\bP}_{C_{1}}{}^{D_{1}}\cdots{\bP}_{C_{n}}{}^{D_{n}}
\left[D_{A}D_{B}T_{D_{1}\cdots D_{n}}-2(\nabla_{A}d) D_{B}T_{D_{1}\cdots D_{n}}\right]\,,\\
&&{\bP}^{AB}P_{C_{1}}{}^{D_{1}}\cdots P_{C_{n}}{}^{D_{n}}
\left[D_{A}D_{B}T_{D_{1}\cdots D_{n}}-2(\nabla_{A}d) D_{B}T_{D_{1}\cdots D_{n}}\right]\,.
\end{eqnarray}

\item Higher order derivatives with gauge covariant vectors, $V_{1},V_{2},\cdots,V_{m}$,  and  a tensor, $T$\,:
\begin{eqnarray}
&&\left(\prod_{i=1}^{m}V_{i}^{B}P_{B}{}^{C}D_{C}\right){\bP}_{A_{1}}{}^{B_{1}}{\bP}_{A_{2}}{}^{B_{2}}\cdots{\bP}_{A_{n}}{}^{B_{n}}T_{B_{1}B_{2}\cdots B_{n}}\,,\\
&&\left(\prod_{i=1}^{m}V_{i}^{B}\bP_{B}{}^{C}D_{C}\right)P_{A_{1}}{}^{B_{1}}P_{A_{2}}{}^{B_{2}}\cdots P_{A_{n}}{}^{B_{n}}T_{B_{1}B_{2}\cdots B_{n}}\,.
\end{eqnarray}

%%%
%%\item Gauge covariant curvatures  with four vector indices:
%%\be
%%\ba{l}
%%\cR_{ABCD}:=?????
%%\ea
%%\ee
%%which are comparable  to the Riemann curvature in ordinary differential geometry.
%%%

\item  Gauge covariant tensor  with two vector indices:
\be
\cR_{AB}:=P_{A}{}^{C}\bP_{B}{}^{D}
\left(S_{CD}+2D_{(C}\nabla_{D)}d\right)\,,\\
\ee
which is  comparable to the Ricci curvature in ordinary differential geometry.

Direct computation, using  (\ref{KHD1}), (\ref{KHD2})\footnote{And also with the help of  the 
computer algebra system, \textit{Cadabra},  developed by Kasper Peeters~\cite{Peeters:2006kp,Peeters:2007wn}.}  shows that the expression inside the bracket, \textit{i.e.~}$S_{AB}+2D_{(A}\nabla_{B)}d$, coincides with ``$\cK_{AB}$" given in  Eq.(4.49) of Ref.\cite{Hohm:2010pp}.  With the decomposition, $\cR_{AB}=\cR_{(AB)}+\cR_{[AB]}$, the symmetric part, $\cR_{(AB)}$, then corresponds to  what was called ``generalized Ricci curvature" in Ref.\cite{Hohm:2010pp}.

\item  Gauge invariant scalar:
\be
\ba{ll}
\cR&:=~~\cH^{AB}\left(4D_{A}\nabla_{B}d-4\nabla_{A}d\,\nabla_{B}d+S_{AB}\right)\,,\\
{}&\,\equiv~~~2P^{AB}\left(4D_{A}\nabla_{B}d-4\nabla_{A}d\,\nabla_{B}d+S_{AB}\right)\,,\\
{}&\,\equiv-2\bP^{AB}\left(4D_{A}\nabla_{B}d-4\nabla_{A}d\,\nabla_{B}d+S_{AB}\right)\,,
\ea
\label{cRscalar}
\ee
which is  comparable to the scalar curvature in ordinary differential geometry. Here, the equivalence relations are due to the level matching constraints  (\ref{constraint1}), (\ref{constraint2}),  implying  the triviality of  the following  quantity, 
\be
4D_{A}\nabla^{A}d-4\nabla_{A}d\,\nabla^{A}d+S\equiv 0\,.
\ee
Direct  computation confirms that the gauge invariant  scalar, $\cR$ in (\ref{cRscalar}), coincides with   $\cR(\cH,d)$ (\ref{cRODD}) first found in \cite{Hohm:2010pp}.  Hence,  after integration by part, we may rewrite the double field theory action (\ref{DFTS}) in a compact form,
\be
\dis{S_{\scriptstyle{\rm{DFT}}}=\int\rd y^{2D}~e^{-2d}\,\cR=\int\rd y^{2D}~e^{-2d}\,\cH^{AB}\left(4\nabla_{A}d\,\nabla_{B}d+S_{AB}\right)\,,}
\ee
of which the  infinitesimal transformation induced by arbitrary $\delta d$ and $\delta P^{AB}$ follows from utilizing (\ref{infPtr}), (\ref{infGtr}), (\ref{infStr}),
\be
\dis{\delta S_{\scriptstyle{\rm{DFT}}}=\int\rd y^{2D}~2e^{-2d}\left[\,\delta P^{AB}\left(S_{AB}+2D_{(A}\nabla_{B)}d\right)
-\delta d\,\cR\,\right]\,.}
\ee
%%%
%%\be
%%\delta\left(e^{-2d}\cR\right)=2e^{-2d}\left[\,\delta P^{AB}\left(S_{AB}+2D_{(A}\nabla_{B)}d\right)-\delta d\,\cR\,\right]
%%\,+\,\mbox{total~derivative}\,.
%%\ee
%%%
This confirms the  result~\cite{Hohm:2010pp} that the equations of motion for the dilaton and  the projection are $\cR=0$ and $\cR_{(AB)}=0$,  respectively. 
\end{itemize}

%%%%%%%%%%%%%%%%%%%%%%%%%%%%%%%%%%%%%%%%%%%%%%%%%%%%%%%%%%%%%%%%%%%%%%%%%%%%%%%%%%%%%%%%%%%%%%%%%%%%%%%%%%%%%%%%%%%%%%%%%%%%%%%%%%%%%%%%%%%%%%%%%%%%%%%%%%%%%%%%%%%%%%%%%%
\section{Comments\label{SecCON}}
%%%%%%%%%%%%%%%%%%%%%%%%%%%%%%%%%%%%%%%%%%%%%%%%%%%%%%%%%%%%%%%%%%%%%%%%%%%%%%%%%%%%%%%%%%%%%%%%%%%%%%%%%%%%%%%%%%%%%%%%%%%%%%%%%%%%%%%%%%%%%%%%%%%%%%%%%%%%%%%%%%%%%%%%%%
In this paper, we have developed a novel differential geometry which is compatible with a $\ODD$ symmetric projection,  in order to analyze  systematically the underlying mathematical structure of  double field theory. We expect our main results, summarized in section \ref{SECsummary},  may  provide useful framework for future  study   on T-duality, such as for the inclusion of fermion, see \textit{e.g.~}\cite{Bergshoeff:1994dg,Bergshoeff:1995as,Berkovits:2008ic,Beisert:2008iq,Godazgar:2010ph}, Ramond sector~\cite{West:2010ev},   non-commutativity~\cite{Lust:2010iy},  the beta functions of double space sigma model~\cite{Berman:2007xn,Berman:2007yf}, 
as well as for the higher derivative corrections to the low energy effective action, (\ref{NSaction}), (\ref{DFTS}), as in \cite{Metsaev:1987zx,Meissner:1996sa,Green:2010sp,Green:2010kv}. The generalization of our  formalism to M-theory is also of interest~\cite{Berman:2010is}.\\

\noindent Since our connection in the projection-compatible derivative is $\sodd$ valued, 
$\Gamma_{CAB}=-\Gamma_{CBA}$, one might wonder whether the global $\ODD$ rotation group is a subgroup of  the gauge group, or whether the T-duality is a gauge symmetry of the double field theory~\cite{Dine:1989vu}.  Yet the answer is negative, as discussed  in \cite{Hohm:2010jy} and also explained further here:   First of all,   to allow for an infinitesimal transformation, we restrict our remaining discussion to the $\SODD$ global rotation.   As for the original discrete T-duality rotation matrix, $\cT$ in (\ref{TODD}), since $\det\cT=(-1)^{D}$, only in the case of $D$ being even, $\cT$ belongs to  $\SODD$. In fact,  as 
\be
{\cT=\left(\ba{cc}0&1\\1&0\ea\right)=\exp\left[\frac{\pi}{2}\left(\ba{cc}\epsilon&-\epsilon\\-\epsilon&\epsilon\ea\right)\right]\,,}
\ee
where $\epsilon$ is a $D\times D$ skew-symmetric matrix of the familiar  form:
\be
\epsilon=\left(\ba{cc}\,0&1\\-1&0\ea\right)\,.
\ee
Generally,     infinitesimal  $\sodd$ parameter is given by a constant matrix, $h_{A}{}^{B}$,  satisfying the anti-symmetric property,  $h_{AB}=-h_{BA}$. Under this  infinitesimal   $\sodd$ rotation, the doubled spacetime  coordinate transforms in  a standard manner,
\be
\ba{ll}
y^{A}~~\longrightarrow~~y^{A}+\delta_{h}y^{A}\,,~~~~&~~~~\delta_{h}y^{A}=y^{B}h_{B}{}^{A}\,,
\ea
\ee
while the transformations of $\cH_{AB}$ and $d$ read, from (\ref{ODDglobal}),
\be
\ba{ll}
\delta_{h}\cH_{AB}=h_{A}{}^{C}\cH_{CB}+h_{B}{}^{C}\cH_{AC}+\delta_{h} y^{C}\partial_{C}\cH_{AB}\,,~~~~&~~~~\delta_{h}d=\delta_{h} y^{C}\partial_{C}d\,.
\ea
\label{sodd}
\ee
Clearly, $\delta_{h}\cH_{AB}\neq\hcL_{{\delta_{h} y}}\cH_{AB}$, and hence   the  $\sodd$ rotation is not a  gauge symmetry.   In fact, since the local gauge parameter should satisfy the Level matching constraint (\ref{XPhi}) along with  other covariant tensors in the theory   which have   dependence only on a $D$-dimensional hyperplane,
\be
\ba{ll}
\partial_{C}X^{A}\partial^{C}T_{B_{1}B_{2}\cdots B_{n}}\equiv 0\,,~~~~&~~~~\partial^{2}T_{B_{1}B_{2}\cdots B_{n}}\equiv 0\,,
\ea
\ee
the gauge symmetry parameter must be  subject to the choice of the hyperplane.  Note that while
\be
\ba{lll}
\partial^{2}\Big(\delta_{h}T_{B_{1}B_{2}\cdots B_{n}}\Big)\equiv 0\,,~~&~~~
\partial^{2}\Big(\hcL_{X} T_{B_{1}B_{2}\cdots B_{n}}\Big)\equiv 0\,,~~&~~~
\partial^{2}\Big(\delta_{h}\hcL_{X}T_{B_{1}B_{2}\cdots B_{n}}\Big)\equiv 0\,,
\ea
\ee
we have 
\be
\partial^{2}\Big(\hcL_{X}\delta_{h}T_{B_{1}B_{2}\cdots B_{n}}\Big)\neq 0\,.
\ee 
Similarly,   while the commutator  between the infinitesimal $\sodd$ rotation and the generalized Lie derivative generating the  local gauge symmetry is again given by a generalized Lie derivative, 
\be
\ba{ll}
{}[\delta_{h},\hcL_{X}]=\hcL_{Y}\,,~~~~&~~~~Y^{A}=y^{B}h_{B}{}^{C}\partial_{C}X^{A}-X^{B}h_{B}{}^{A}=\delta_{h}X^{A}\,,
\ea
\ee
generically $\hcL_{Y}$ does not generate any symmetry of the   double field theory, as the parameter, $Y^{A}$, does not necessarily  meet the level matching constraint (\ref{constraint2}),
\be
\partial^{C}Y^{A}\partial_{C}T_{B_{1}B_{2}\cdots B_{n}} \equiv 
h^{CD}\partial_{D}X^{A}\partial_{C}T_{B_{1}B_{2}\cdots B_{n}} \neq 0\,.
\ee
To summarize,  as the $\ODD$ transformation  rotates   the entire  hyperplane on which the double field theory fields live,  it  corresponds to   \textit{a priori}  a duality, and only after  dimensional reduction is taken   it may become   a symmetry of the action.\\

We conclude  by commenting that,  in double field theory, the differential geometry with a  $\ODD$ symmetric  projection on a `flat' doubled spacetime  contains  intriguingly the Riemann geometry on the `curved' $D$-dimensional  hyperplane.

~\\
~\\
~\\
\noindent\textbf{Acknowledgements}\\
We thank the organizers of the International School On
Strings And Fundamental Physics,  M\"{u}nchen (July 25 -- August 6, 2010),   where
this work was initiated thanks to  the lecture by   Barton Zwiebach.  The work was supported by the National Research Foundation of Korea(NRF) grants  funded by the Korea government(MEST) with the grant numbers 2005-0049409 (CQUeST)  and 2010-0002980.
\newpage

%%%%%%%%%%%%%%%%%%%%%%%%%%%%%%%%%%%%%%%%%%%%%%%%%%%%%%%%%%%%%%%%%%%%%%%%%%%%%%%%%%%%%%%%%
%%%%%%%%%%%%%%%%%%%%%%%%%%%%%%%%%%%%%%%%%%%%%%%%%%%%%%%%%%%%%%%%%%%%%%%%%%%%%%%%%%%%%%%%%
%%
%%
\appendix

%%%%%%%%%%%%%%%%%%%%%%%%%%%%%%%%%%%%%%%%%%%%%%%%%%%%%%%%%%%%%%%%%%%%%%%%%%%%%%%%%%%%%%%%%%%%%%%%%%%%%%%%%%%%%%%%%%%%%%%%%%%%%%%%%%%%%%%%%%%%%%%%%%%%%%%%%%%%%%%%%%%%%%%%%%%%
\section{Useful relations\label{SECAPPENDIXuseful}}
Here we write some useful identities. Under the gauge  transformation of a tensor (\ref{covtensor}), the ordinary derivative of a generic tensor, $T_{A_{1}A_{2}\cdots A_{n}}$, transforms as
\be
(\delta_{X}-\hcL_{X})\partial_{C}T_{A_{1}A_{2}\cdots A_{n}}=\partial_{B}X_{C}\partial^{B}T_{A_{1}A_{2}\cdots A_{n}}+\sum_{i=1}^{ㅜ}(\partial_{C}\partial_{A_{i}}X^{B}-\partial_{C}\partial^{B}X_{A_{i}})T_{A_{1}\cdots A_{i-1}BA_{i+1}\cdots A_{n}}\,.
\ee
Especially for the projection, $P$, we have 
\be
(\delta_{X}-\hcL_{X})\partial_{C}P_{DE}=\partial_{B}X_{C}\partial^{B}P_{DE}+(\partial_{C}\partial_{D}X^{B}-\partial_{C}\partial^{B}X_{D})P_{BE}+
(\partial_{C}\partial_{E}X^{B}-\partial_{C}\partial^{B}X_{E})P_{DB}\,,
\ee
such that 
\be
\ba{l}
(\delta_{X}-\hcL_{X})(P\partial_{C}P)_{DE}=\partial_{B}X_{C}(P\partial^{B}P)_{DE}+
2\partial_{C}\partial_{[A}X_{B]}P_{D}{}^{B}{\bP}_{E}{}^{A}\,,\\
(\delta_{X}-\hcL_{X})(\partial_{C}PP)_{DE}=\partial_{B}X_{C}(\partial^{B}PP)_{DE}
-2\partial_{C}\partial_{[A}X_{B]}{\bP}_{D}{}^{B}P_{E}{}^{A}\,,
\ea
\ee
and
\be
\ba{l}
(\delta_{X}-\hcL_{X})P_{C}{}^{F}(P\partial_{F}P)_{DE}
=P_{C}{}^{F}\partial_{B}X_{F}(P\partial^{B}P)_{DE}-2
P_{C}{}^{F}P_{D}{}^{A}{\bP}_{E}{}^{B}\partial_{F}\partial_{[A}X_{B]}\,,\\
(\delta_{X}-\hcL_{X}){\bP}_{C}{}^{F}(\partial_{F}PP)_{DE}
={\bP}_{C}{}^{F}\partial_{B}X_{F}(\partial^{B}PP)_{DE}
+2{\bP}_{C}{}^{F}{\bP}_{D}{}^{A}P_{E}{}^{B}\partial_{F}\partial_{[A}X_{B]}\,,\\
(\delta_{X}-\hcL_{X})\left[\partial_{C}P(P-\bP)\right]_{DE}
=\partial_{B}X_{C}\left[\partial^{B}P(P-\bP)\right]_{DE}
+2\left[P_{D}{}^{A}{\bP}_{E}{}^{B}+{\bP}_{D}{}^{A}P_{E}{}^{B}\right]
\partial_{C}\partial_{[A}X_{B]}\,.
\ea
\label{PPPtr}
\ee
In particular,  the gauge transformations of (\ref{chiralcandi}) and (\ref{antichiralcandi}) are
\be
\ba{l}
(\delta_{X}-\hcL_{X})P_{D[A}(P\partial^{D}P)_{B]C}=\partial_{D}X^{E}P_{E[A}(P\partial^{D}P)_{B]C}+
 P_{A}{}^{D}P_{B}{}^{E}{\bP}_{C}{}^{F}\partial_{F}\partial_{[D}X_{E]}\,,\\
 (\delta_{X}-\hcL_{X}){\bP}_{D[A}(\partial^{D}PP)_{B]C} =\partial_{D}X^{E}{\bP}_{E[A}(\partial^{D}PP)_{B]C}-
 {\bP}_{A}{}^{D}{\bP}_{B}{}^{E}P_{C}{}^{F}\partial_{F}\partial_{[D}X_{E]}\,.
\ea
\label{ForG}
\ee
For the connection (\ref{connectionG}), we have
\be
\ba{l}
(\delta_{X}-\hcL_{X})\Gamma_{CAB}\\
=\partial_{D}X_{E}\left[
-\delta^{E}{}_{C}[\partial^{D}P(P-\bP)]_{AB}-2P^{E}{}_{[A}(P\partial^{D}P)_{B]C}+2{\bP}^{E}{}_{[A}(\partial^{D}PP)_{B]C}\right]\\
~~~~~-\left(\delta_{A}{}^{D}P_{B}{}^{E}{\bP}_{C}{}^{F}+P_{A}{}^{D}{\bP}_{B}{}^{E}\delta_{C}{}^{F}+
{\bP}_{A}{}^{D}\delta_{B}{}^{E}P_{C}{}^{F}\right)(\partial_{F}\partial_{D}X_{E}-\partial_{F}\partial_{E}X_{D})\\
=\partial_{D}X_{E}\left[
-\delta^{E}{}_{C}[\partial^{D}P(P-\bP)]_{AB}-2P^{E}{}_{[A}\partial^{D}P_{B]C}+2\delta^{E}{}_{[A}(\partial^{D}PP)_{B]C}\right]\\
~~~~~+\left(P_{A}{}^{D}P_{B}{}^{E}P_{C}{}^{F}+
{\bP}_{A}{}^{D}{\bP}_{B}{}^{E}{\bP}_{C}{}^{F}-\delta_{A}{}^{D}\delta_{B}{}^{E}\delta_{C}{}^{F}\right)(\partial_{F}\partial_{D}X_{E}-\partial_{F}\partial_{E}X_{D})\\
\equiv \left(P_{A}{}^{D}P_{B}{}^{E}P_{C}{}^{F}+
{\bP}_{A}{}^{D}{\bP}_{B}{}^{E}{\bP}_{C}{}^{F}-\delta_{A}{}^{D}\delta_{B}{}^{E}\delta_{C}{}^{F}\right)(\partial_{F}\partial_{D}X_{E}-\partial_{F}\partial_{E}X_{D})\,.
\ea
%\label{Godd}
\ee
Especially, up to the constraint, we get
\be
\ba{l}
(\delta_{X}-\hcL_{X})\Gamma^{B}{}_{BA}\equiv\partial_{A}\partial_{B}X^{B}+\left(P_{A}{}^{B}P^{CD}+{\bP}_{A}{}^{B}{\bP}^{CD}\right)2\partial_{C}\partial_{[D}X_{B]}\,,\\
P_{A}{}^{D}P_{B}{}^{E}(\delta_{X}-\hcL_{X})\Gamma_{CDE}\equiv-2P_{A}{}^{D}P_{B}{}^{E}\bP_{C}{}^{F}\partial_{F}\partial_{[D}X_{E]}\,,\\
P_{C}{}^{F}P_{A}{}^{D}(\delta_{X}-\hcL_{X})\Gamma_{FDB}\equiv-2P_{C}{}^{F}P_{A}{}^{D}\bP_{B}{}^{E}\partial_{F}\partial_{[D}X_{E]}\,.
\ea
\ee
For the projection-compatible derivative of a generic tensor, we have
\be
\ba{l}
\left(\delta_{X}-\hcL_{X}\right)D_{C}T_{A_{1}A_{2}\cdots A_{n}}\\
=\partial^{D}X_{C}\nD_{D}T_{A_{1}A_{2}\cdots A_{n}}+2\sum_{i}\partial_{D}X_{E}\left[P^{E}{}_{[B}\partial^{D}P_{A_{i}]C}
+\delta^{E}{}_{[A_{i}}(\partial^{D}PP)_{B]C}\right]T_{A_{1}\cdots A_{i-1}}{}^{B}{}_{A_{i+1}\cdots A_{n}}\\
~~~~+2\sum_{i}\left(P_{A_{i}}{}^{D}P_{B}{}^{E}P_{C}{}^{F}+{\bP}_{A_{i}}{}^{D}{\bP}_{B}{}^{E}{\bP}_{C}{}^{F}\right)
\partial_{F}\partial_{[D}X_{E]}T_{A_{1}\cdots A_{i-1}}{}^{B}{}_{A_{i+1}\cdots A_{n}}\\
\equiv 2\sum_{i}\left(P_{A_{i}}{}^{D}P_{B}{}^{E}P_{C}{}^{F}+{\bP}_{A_{i}}{}^{D}{\bP}_{B}{}^{E}{\bP}_{C}{}^{F}\right)
\partial_{F}\partial_{[D}X_{E]}T_{A_{1}\cdots A_{i-1}}{}^{B}{}_{A_{i+1}\cdots A_{n}}\,.
\ea
%\label{DTodd}
\ee
From (\ref{RABsym}), we obtain
\be
(\delta_{X}-\hcL_{X})\left(P_{A}{}^{C}\bP_{B}{}^{D}S_{CD}\right)
\equiv -\left(P^{EF}P_{A}{}^{G}\bP_{B}{}^{C}D_{C}+\bP^{EF}\bP_{B}{}^{G}P_{A}{}^{C}
D_{C}\right)\partial_{E}\partial_{[F}X_{G]}\,.
\ee
From the symmetric properties of the connection (\ref{sympropG}), we get
\be
\Gamma_{ABC}\Gamma^{ABC}=2\Gamma_{ABC}\Gamma^{BAC}\,.
\label{Gsquare}
\ee
In order to see  that our results agree with  Ref.\cite{Hohm:2010pp}, it is useful to write explicitly,
\be
\textstyle{\Gamma_{CAB} = \frac{1}{4} \cH_{B}{}^{D} \partial_{D} \cH_{A C} + \frac{1}{4} \cH_{C}{}^{D} \partial_{A} \cH_{DB} +\half \cH_{A}{}^{D}\partial_{C}\cH_{DB}
 -\frac{1}{4} \cH_{A}{}^{D}\partial_{D} \cH_{BC}   -\frac{1}{4}\cH_{C}{}^{D}\partial_{B}\cH_{DA}\,,}
\label{KHD1}
\ee
and
\be
\ba{l}
P_{A}{}^{C} \bP_{B}{}^{D}\Gamma^{EF}{}_{(C}\Gamma_{D)EF} \equiv -\frac{1}{8}
P_{A}{}^{C} \bP_{B}{}^{D}\partial_{C}\cH^{EF}\partial_{D}\cH_{EF}\,,\\
P_{A}{}^{C} \bP_{B}{}^{D}\Gamma_{EF(C}\Gamma^{EF}{}_{D)} \equiv
P_{A}{}^{C} \bP_{B}{}^{D}\left( \half 
\partial_{E}\cH_{F(C}\partial_{D)}\cH^{EF}- \half \partial_{E}\cH_{FC}\partial^{F}\cH^{E}{}_{D}\right)\,,\\
P_{A}{}^{C} \bP_{B}{}^{D}\Gamma_{EF(C}\Gamma^{FE}{}_{D)} \equiv P_{A}{}^{C} \bP_{B}{}^{D}\left(\half\partial_{E}\cH_{F(C}\partial_{D)}\cH^{EF}
-\half \partial_{E}\cH_{FC} \partial^{F}\cH^{E}{}_{D}-\frac{1}{8}\partial_{C}\cH^{EF}\partial_{D}\cH_{EF}\right)\,.
\ea
\label{KHD2}
\ee
\newpage

%%%%%%%%%%%%%%%%%%%%%%%%%%%%%%%%%%%%%%%%%%%%%%%%%%%%%%%%%%%%%%%%%%%%%%%%%%%%%%%%%%%%%%%%%%%%%%%%%%%%%%%%%%%%%%%%%%%%%%%%%%%%%%%%%%%%%%%%%%%%%%%%%%%%%%%%%%%%%%%%%%%%%%%%%%

\end{document}